\documentclass[english]{article}
\usepackage[T1]{fontenc}
\usepackage[latin9]{inputenc}
\usepackage{color}

\makeatletter
\newcommand{\lyxaddress}[1]{
\par {\raggedright #1
\vspace{1.4em}
\noindent\par}
}

\makeatother

\usepackage{babel}
\begin{document}

\title{\textbf{Comment on Eur. Phys. J. Plus 133, 261 (2018) by Kholmetskii
et al. (The YARK theory of gravity is wrong)}}

\author{\textbf{Christian Corda}}
\maketitle

\lyxaddress{\begin{center}
International Institute for Applicable Mathematics and Information
Sciences\textit{, }Adarshnagar, Hyderabad 500063 (India)\textit{\emph{.}}\textit{ }
\par\end{center}}

\lyxaddress{\begin{center}
\textit{E-mail address:} \textcolor{blue}{cordac.galilei@gmail.com} 
\par\end{center}}
\begin{abstract}
We show that the YARK theory of gravity, proposed in Eur. Phys. J.
Plus 133, 261 (2018) by Kholmetskii et al. and in previous papers
of the same research group is wrong.
\end{abstract}
\begin{quotation}
\end{quotation}
Despite it is well known that completely non-metric gravitational
theories macroscopically violate Einstein's equivalence principle
(EEP), in an astonishing way, various papers on a completely non-metric
gravitational theory, today self-called \emph{the YARK theory of gravitation}
from the initials of the proper surnames of its authors, have been
recently published in various journals {[}1 - 15{]}, included Eur.
Phys. J. Plus {[}1 - 3{]}. We recall that the YARK theory of gravitation
has been originally proposed by T. Yarman in Foundation of Physics
\cite{key-10}. This happened in the well known period of time in
which Foundation of Physics published a lot of wrong and non-standard
results, before G. 't Hooft's management \cite{key-11}. After that,
various papers on the YARK theory of gravitation have been published
by T. Yarman and collaborators (O. Yarman, A. L. Kholmetskii and M.
Arik. Hereafter we will refer to them as the YARK group) {[}1 - 9,
11 - 15{]}. In all the cited papers, the YARK group also insinuated
that the YARK theory of gravitation should replace Einstein's GTR
and that the GTR has various problems {[}1 - 15{]}. In addition, the
YARK theory of gravitation should be in agreement with various experiments
on earth and astrophysical observations {[}1 - 15{]}. 

Let us start our discussion by recalling that EEP has today a strong,
unchallengeable empiric evidence \cite{key-17}. The weak equivalence
principle (WEP), which is included in the EEP, states that the mass
of the body is proportional to its weight \cite{key-17}, or, alternatively,
that the trajectory of a freely falling test mass (i.e. a mass which
is not acted upon by such forces as electromagnetism and too small
to be affected by tidal gravitational forces) is independent of the
mass internal structure and composition \cite{key-17}. The WEP also
states the \emph{Universality of Free Fall}, which means that all
the bodies fall with the same acceleration \cite{key-17}. The EEP
is a more powerful concept stating that \cite{key-17}: 

a) WEP is valid; 

b) the outcome of any local non-gravitational experiment is independent
of the velocity of the freely-falling reference frame in which such
an experiment is performed (local Lorentz invariance, LLI); 

c) the outcome of any local non-gravitational experiment is independent
of where and when in the universe such an experiment is performed
(local position invariance, LPI). 

Thus, it is a very natural and intuitive statement that if EEP is
valid, then gravitation must be a \textquotedblleft curved space-time\textquotedblright{}
phenomenon \cite{key-17}. This means that the effects of gravitation
are completely equivalent to the effects of living in a curved space-time
\cite{key-17}. In other words, gravity is not a force. Instead, it
is inertia in a curved space-time manifold \cite{key-18}. Thus, one
sees that, if EEP is valid, then in local freely falling frames, one
needs the laws governing experiments to be independent of the velocity
of the frame (LLI), with constant values for the various atomic constants
(in order to guarantee LPI) \cite{key-17}. The only laws of Nature
that fulfill this are the ones being compatible with the special theory
of relativity, such as Maxwell\textquoteright s equations of electromagnetism,
and the standard model of particles \cite{key-17}. In addition, in
a local freely falling frame, test masses appear to be not accelerated,
and then moving on straight lines \cite{key-17}. Such \emph{locally
straight} lines obviously correspond to \emph{geodesics} in a curved
space-time \cite{key-17}. The strong, unchallengeable consequence
of this argument is that the only viable theories of gravity are the
metric theories of gravity, or possibly theories that are metric apart
from very weak or short-range non-metric couplings \cite{key-17,key-18}.
We stress that there is a rigorous mathematical demonstration of our
last statement. Let us assume:
\begin{enumerate}
\item The existence of a continuous space-time manifold.
\item The validity of EEP.
\end{enumerate}
Then, following \cite{key-19,key-20}, one supposes that no particles
are accelerating in the neighborhood of a point-event with respect
to a freely falling coordinate system $\left(X^{\mu}\right)$. Setting
$T=X^{0}$ we can write \cite{key-19,key-20},

\begin{equation}
\frac{d^{2}X^{\mu}}{dT^{2}}=0,\label{eq: free fall}
\end{equation}
which is locally applicable in free fall. Now, the chain rule gives
\cite{key-19,key-20}

\begin{equation}
\frac{dX^{\mu}}{dT}=\frac{dx^{\nu}}{dT}\frac{\partial X^{\mu}}{\partial x^{\nu}}.\label{eq: chain rule}
\end{equation}
If we differentiate eq. (\ref{eq: chain rule}) with respect to $T$
we get \cite{key-19,key-20}
\begin{equation}
\frac{d^{2}X^{\mu}}{dT^{2}}=\frac{d^{2}x^{\nu}}{dT^{2}}\frac{\partial X^{\mu}}{\partial x^{\nu}}+\frac{dx^{\nu}}{dT}\frac{dx^{\alpha}}{dT}\frac{\partial^{2}X^{\mu}}{\partial x^{\nu}\partial x^{\alpha}}.\label{eq: Differentiating}
\end{equation}
Let us combine eqs. (\ref{eq: free fall}) and (\ref{eq: Differentiating}).
Then we obtain \cite{key-19,key-20}
\begin{equation}
\frac{d^{2}x^{\nu}}{dT^{2}}\frac{\partial X^{\mu}}{\partial x^{\nu}}=-\frac{dx^{\nu}}{dT}\frac{dx^{\alpha}}{dT}\frac{\partial^{2}X^{\mu}}{\partial x^{\nu}\partial x^{\alpha}}.\label{eq: nullo}
\end{equation}
If one multiplies both sides of eq. (\ref{eq: nullo}) by $\frac{\partial x^{\lambda}}{\partial X^{\mu}}$
one obtains \cite{key-19,key-20}

\begin{equation}
\frac{d^{2}x^{\lambda}}{dT^{2}}=-\frac{dx^{\nu}}{dT}\frac{dx^{\alpha}}{dT}\left[\frac{\partial^{2}X^{\mu}}{\partial x^{\nu}\partial x^{\alpha}}\frac{\partial x^{\lambda}}{\partial X^{\mu}}\right].\label{eq: Multiplying}
\end{equation}
By putting $t=x^{0}$ and by using again the chain rule, one can eliminate
$T$ in favor of the coordinate time $t$ obtaining \cite{key-19,key-20}
\begin{equation}
\frac{d^{2}x^{\lambda}}{dt^{2}}=-\frac{dx^{\nu}}{dt}\frac{dx^{\alpha}}{dt}\left[\frac{\partial^{2}X^{\mu}}{\partial x^{\nu}\partial x^{\alpha}}\frac{\partial x^{\lambda}}{\partial X^{\mu}}\right]+\frac{dx^{\nu}}{dt}\frac{dx^{\alpha}}{dt}\frac{dx^{\lambda}}{dt}\left[\frac{\partial^{2}X^{\mu}}{\partial x^{\nu}\partial x^{\alpha}}\frac{\partial x^{0}}{\partial X^{\mu}}\right].\label{eq: quasi geodetiche}
\end{equation}
We recall that the bracketed terms involving the relationship between
local coordinates $X$ and general coordinates $x$ are functions
of the general coordinates \cite{key-19,key-20}. In that way, eq.
(\ref{eq: quasi geodetiche}) gives immediately the geodesic equation
of motion using the coordinate time $t$ as parameter \cite{key-19,key-20}
\begin{equation}
\frac{d^{2}x^{\lambda}}{dt^{2}}=-\Gamma_{\nu\alpha}^{\lambda}\frac{dx^{\nu}}{dt}\frac{dx^{\alpha}}{dt}+\Gamma_{\nu\alpha}^{0}\frac{dx^{\nu}}{dt}\frac{dx^{\alpha}}{dt}\frac{dx^{\lambda}}{dt},\label{eq: geodetiche rispetto a t}
\end{equation}
which can be re-written in terms of the scalar parameter $s$ as the
standard geodesic equation \cite{key-19,key-20}
\begin{equation}
\frac{d^{2}x^{\lambda}}{ds^{2}}=-\Gamma_{\nu\alpha}^{\lambda}\frac{dx^{\nu}}{ds}\frac{dx^{\alpha}}{ds}.\label{eq: geodetiche rispetto ad s}
\end{equation}
Thus, we have shown that the two assumptions of the existence of a
space-time manifold and of the validity of EEP \textbf{rigorously
imply that the gravitational motion must be geodesics}. In other words,
the correct gravitational theory \textbf{must be a metric theory}
(or a possibly theory that is metric apart from very weak or short-range
non-metric couplings \cite{key-17}, but this is NOT the case of YARK
theory). We stress that the YARK group did not understand this key
point in \cite{key-5}. In fact, in \cite{key-5} they verbatim claim
that ``said derivation (i.e. the above one) is exclusively restricted
to the domain of a purely metric theory''. This is incorrect. We
indeed did NOT assume that the gravitational theory must be metric.
We assumed ONLY the existence of a continuous space-time manifold
and the validity of EEP. Through our rigorous mathematical computation
we have shown that these two assumptions imply that the gravitational
theory must be purely metric. In other words, this was a conclusion
and a result. It was NOT an assumption, contrary to the claims of
the YARK group in \cite{key-5}. In addition, in \cite{key-5} the
YARK group generated further confusion by verbatim adding that ``in
YARK theory the derivatives $\frac{\partial X^{\mu}}{\partial x^{\nu}}$
already do not depend explicitly on spatial coordinates, but only
on the static gravitational binding energy''. This is another basic
mistake which is connected with the issue that the YARK group claims
that YARK theory permits to localize the gravitational energy {[}1
- 15{]}. In the opinion of the YARK group the gravitational energy
should remain a non-vanishing quantity in all plausible frames of
reference {[}1 - 15{]}. This should permit to write down, explicitly,
a stress-energy tensor for the gravitational field {[}1 - 15{]}. Clearly,
the YARK group does not understand the real meaning of EEP. In fact,
another consequence of EEP is that one can always find in any given
locality a reference's frame (the local Lorentz reference's frame)
in which ALL local gravitational fields are null. No local gravitational
fields means no local gravitational energy-momentum and, in turn,
no stress-energy tensor for the gravitational field \cite{key-18}.
Also in this case, the YARK group claims that this statement is again
strictly applicable only to metric theories, as is the case with the
GTR \cite{key-5}. This is again wrong. In fact, it is well known
that this is a mere consequence of Einstein's 'happiest thought' that
a freely falling body has not weight \cite{key-21}. Einstein's 'happiest
thought' is indeed at the foundation of both of WEP and EEP. In other
words, EEP has two rigorous consequences:\medskip{}

{*} Gravitational motion must be geodesic.

{*}{*} The gravitational energy cannot be localized. \medskip{}

Both of points {*} and {*}{*} are consequences of EEP and, in turn,
one does NOT need the assumption that a gravitational theory must
be metric to verify points {*} and {*}{*}. The metric behavior of
a gravitational theory is \textbf{a consequence} of point {*} instead
of an a priori assumption. 

Clearly, based on the extreme precision on which the EEP is today
tested and verified \cite{key-17}, the demonstration that we have
reviewed above - i.e. that geodesic motions arise from the EEP - \textbf{ultimately
rules out YARK theory}. In fact, that theory is founded on the absence
of curvature {[}1 - 15{]} and so has a non-viable behavior. In other
words, it is wrong. Despite the claims of the YARK group that the
YARK theory of gravitation should be in agreement with various experiments
on earth and astrophysical observations {[}1 - 15{]} (but in the following
we will show that the YARK group is basically wrong in its YARK interpretation
of the first gravitational wave signal detected by LIGO \cite{key-23}),
the YARK theory of gravitation is indeed in macroscopic contrast with
the strongest observational constrain that a gravitational theory
must satisfy, that is the EEP, which is founded on tons of experimental
data \cite{key-17}. Recently, we discussed this issue in some private
communications with the leader theorist of the YARK group, i.e. T.
Yarman \cite{key-24}. T. Yarman honestly admitted that the above
derivation is correct, but he now claims that it is the EEP which,
in his opinion, does not work \cite{key-24}. T. Yarman's opposition
to the EEP is the following. He claims the the ratio between the gravitational
mass and the inertial mass, i.e. \cite{key-24}
\begin{equation}
\frac{m_{g}}{m_{i}}\equiv K\label{eq: K}
\end{equation}
is not universal, despite the quantity (\ref{eq: K}) is today tested
with the enormous precision of 1 part in $10^{14}$ by experiments
\cite{key-25,key-26}. T. Yarman claims indeed that all the experiments
concerning the ratio between the gravitational mass and the inertial
mass, starting from the historical experiments of Eötvös \cite{key-27}
and the subsequent \cite{key-28,key-29} till the most recent ones
\cite{key-25,key-26}, have been misinterpreted by the scientific
community \cite{key-24}. He thinks that the experiments {[}25 - 29{]},
and other ones, permit only to test the proportionality between the
gravitational mass and the inertial mass rather than testing their
effective equivalence \cite{key-24}. In other words, T. Yarman claims
that, in order to adopt $K=1$ in Eq. (\ref{eq: K}), one has further
to assume that $K$ is a universal constant, without yet any rigorous
experimental ground behind it \cite{key-24}. In addition, he claims
to have shown that it is instead \cite{key-24}
\begin{equation}
K=1-\frac{v^{2}}{c^{2}},\label{eq: K sbagliato}
\end{equation}
being $v$ the velocity of the mass with respect to the chosen reference
frame and $c$ the speed of light. In fact, T. Yarman claims to have
shown that the gravitational mass is given by \cite{key-24} 
\begin{equation}
m_{g}=m_{0}exp(-\alpha)\sqrt{1-\frac{v^{2}}{c^{2}}}\label{eq: mg sbagliata}
\end{equation}
and that the inertial mass is given by 
\begin{equation}
m_{i}=m_{0}\frac{exp(-\alpha)}{\sqrt{1-\frac{v^{2}}{c^{2}}}},\label{eq: mi sbagliata}
\end{equation}
where $m_{0}$ is the rest mass of the object at hand, in free space,
and $\alpha=\frac{GM}{rc^{2}},$ being $M$ the source of the gravitational
field, $G$ the Newtonian gravitational constant and $r$ the distance
of the moving mass from the source of the gravitational field. We
immediately recognize that $\gamma\equiv\frac{1}{\sqrt{1-\frac{v^{2}}{c^{2}}}}$
is the well known Lorentz factor of the special theory of relativity
(STR) \cite{key-30}. Thus, setting $M=0$ in Eq. (\ref{eq: mi sbagliata}),
one immediately gets 
\begin{equation}
m_{i}=m_{0}\frac{1}{\sqrt{1-\frac{v^{2}}{c^{2}}}}\label{eq: mi giusta}
\end{equation}
which is the traditional inertial mass of the STR \cite{key-30}.
In any case, even assuming that T. Yarman is correct (and we think
that he is not), he misses a fundamental point here. The EEP has \emph{local}
behavior. In a local Lorentz frame it is $v=0$ and one immediately
obtains $K=1$ in Eqs. (\ref{eq: K}) and (\ref{eq: K sbagliato}).
In other words, a free falling observed cannot distinguish between
gravitational motion and inertial motion even assuming that T. Yarman's
analysis in \cite{key-24} is correct. Therefore Einstein's 'happiest
thought' \cite{key-21} is ultimately confirmed also in this case.
Hence, T. Yarman's analysis in \cite{key-24} cannot invalidate the
above rigorous analysis that the EEP implies the geodesic motion (\ref{eq: geodetiche rispetto ad s})
which, in turn, ultimately rules out the YARK theory of gravitation.
It is also useful adding the following considerations. As it is not
Lorentz invariant, within the STR the concept of inertial mass is
dubious and/or ambiguous. Despite such difficulties (to attribute
a value to inertial mass in the STR), we think that it still remains
a description of the resistance of an object to change its proper
velocity. Such a resistance becomes dependent on the frame of reference,
see Eq. (\ref{eq: mi giusta}). On one hand, this makes the concept
of inertial mass of little practical use in the STR. On the other
hand, it does not damage the underlying idea of what it really represents.

Finally, we discuss the claims of the YARK group that the YARK theory
can predict everything what is predicted through LIGO signals \cite{key-9}.
Actually, in \cite{key-9} we find that in the the YARK theory the
two arms are altered in the same way (and not depending on the direction
of the incoming perturbation). This is well known to be an elementary
mistake. In fact, the LIGO signal is given by the phase difference
between the two interferometer's arms, see for example \cite{key-31}.
Thus, if the two arms are altered in the same way (and not depending
on the direction of the incoming perturbation) LIGO would not produce
any output. This is a further proof that the YARK theory of gravity
is wrong.

\subsection*{Conclusion remarks}

In this Comment we have shown that the YARK theory of gravity, proposed
in {[}1 - 15{]} is wrong for two important reasons. First, it is not
consistent with the EEP; second, contrary to the claims in \cite{key-9},
it cannot predict everything what is predicted through LIGO signals.
We hope that this Comment will prevent future publications on the
wrong YARK theory of gravity in Eur. Phys. J. Plus and in other journals.

\subsection*{Acknowledgements }

It is a pleasure to thank the Editor in Chief of Eur. Phys. J. Plus,
Prof. Paolo Biscari, for inviting us to write this Comment. The author
also thanks Mr. T. Yarman for private discussions on the issues of
this Comment and an unknown referee for useful comments.

\end{document}